# Carbon mono and dioxide hydrogenation over pure and metal oxide decorated graphene oxide substrates: insight from DFT

Danil W. Boukhvalov

School of Computational Studies, Korea Institute for Advanced Study (KIAS), Seoul 130-722, Korea.

*Based on first principles density functional theory calculations we explore the energetics of the conversion of carbon mono and dioxide to methane over graphene oxide surfaces. Similar to the recently discovered hydration of various organic species over this catalyst, the transfer of hydrogen atoms from hydroxyl groups of graphene oxide provide a step by step transformation hydrogenation of carbon oxides. Estimated yields of modeled reactions at room temperature are about 0.01% for the carbon mono and dioxide. For the modeling of graphene oxide/metal oxide composites, calculations in the presence of $MO_2$ (where M = V, Cr, Mn, Fe) have been performed. Results of these calculations demonstrate significant decreases of the energy costs and increases of reaction yields to 0.07 %, which is comparable to the efficiency of these reactions over platinum and ruthenium-based photocatalysts. Increasing the temperature to the value 100 °C should provide the total conversion of carbon mono and dioxides.*

E-mail: danil@kias.re.kr

**Introduction**

Carbon-based systems have been discussed as prospective alternatives for conventional metal-based catalysts over the past decade. These studies were motivated by the abundance, low cost, light weight and diversity of structural allotropes of carbon. [1] The discovery and rapid progress of large scale and low cost production of two dimensional carbon (graphene) [2,3] makes this material attractive for the various applications and encourages the exploration of catalytic properties of graphene derivatives. Another advantage of carbon-based catalysts is the absence of the surface poisoning problem leading to a significant increase in the number of the working cycles compared to the conventional metal-based catalysts.[4] For catalytic applications, the most important differences between graphene and carbon nanotubes or porous carbons is (i) a larger surface/weight ratio, (ii) higher accessibility of the surface and (iii) more controllable chemical composition (negligible quantity of contaminations and graphitized or defected areas).

The first experimental reports which considered the catalytic activity of graphene demonstrated the efficiency of nitrogen doped graphene for oxygen reduction reactions [4-9] and graphene oxide (GO) for hydration and oxidation of various organic compounds. [10,11] The multiplicity of possible routes of the graphene surface decoration and doping by nitrogen, boron and sulfur [4-9, 12-14] and the simplicity of large scale production of GO necessitate further studies of the catalytic properties of modified graphene. Our recent modeling of the oxygen reduction reaction over nitrogen doped graphene [15] and the oxidation and hydration of various organic species on graphene oxide substrate [16] using density-functional theory (DFT) methods demonstrate the ability of DFT for the description of both the catalytic properties of pure and modified graphene and the different natures of the catalytic properties of doped graphene and GO.

Hydrogenation of carbon dioxide and the conversion of carbon monoxide is the subject of numerous works (see Refs. 17-26 and references therein). In these studies, various catalysts and mechanisms of reactions have been proposed for achieving this goal. The main difficulties of these reactions using the studied catalysts is a very low output, corresponding with colossal energy barriers of the intermediate steps and problems with capturing the gases over or inside catalysts, employment of rare elements [21-26] or the requirement of the additional molecular hydrogen. [25-26] The impenetrability of graphene oxide paper for various gases [27] and surprisingly high yields of various reactions over GO substrates at moderate temperatures, [10,11] and resent report about photocatalytic conversion of CO2 to methanol [28] encourage the modeling of the reduction of carbon oxides in the presence of a GO catalyst. In this article we report the results of modeling carbon mono and dioxide transformations to methane over the surface of reduced GO (both pristine and functionalized by transitional metal oxides).

**Computational method**

The modeling for this work was performed using density functional theory (DFT) in the pseudopotential code SIESTA, [29] as was done in our previous studies.[15,16,30,31] All calculations were performed using the generalized gradient approximation (GGA-PBE) [32] with spin-polarization. Full optimization of the atomic positions was performed. During the optimization, the ion cores are described by norm-conserving non-relativistic pseudo-potentials [33] with cut off radii 1.14, 1.45 and 1.25 a.u. for C, O and H respectively, and the wavefunctions are expanded with a double-ζ plus polarization basis of localized orbitals for carbon and oxygen, and a double-ζ basis for hydrogen. Optimizations of the force and total energy were performed with a accuracies of 0.04 eV/Å and 1 meV, respectively. All calculations were carried out with an energy mesh cut-off of 360 Ry and a k-point mesh of 8×6×1 in the Monkhorst-Park scheme. [34]

**Modeling the catalytic process**

In our previous model of the reduction of different organic species over GO substrates [16] we proposed a mechanism of reaction: hydrogen from the hydroxyl groups of GO migrate to the reducible species and the residual oxygen atom from the hydroxyl group transforms to an epoxy group on the carbon surface (see Fig. 1 a-c). Recent experimental works are also demonstrated instability of graphene oxide chemical structure at the temperatures about 100 °C caused migration of the functional groups to substrate [35] or interlayer space. [36] The catalytic process provides the reduction of GO to a significant level [37] with further re-oxidation from the environment (air, water). [10, 16] For modeling the probable pathway of carbon dioxide reduction we use the model proposed for the transformation of this species to methane over a ZnO surface:

$CO_2 + 2H^+ + 2e^- \rightarrow HCOOH$ (1)

$HCOOH + 2H^+ + 2e^- \rightarrow HCHO + H_2O$ (2)

$HCHO + 2H^+ + 2e^- \rightarrow CH_3OH$ (3)

$CH_3OH + 2H^+ + 2e^- \rightarrow CH_4 + H_2O$ (4) [19,20]

This mechanism of reaction has been chosen because it requires a minimal number of reactions of GO with molecules, leading to higher reaction yields, and it does not require additional molecular hydrogen because hydroxyl groups of GO is the source of hydrogen atoms.[16] Employment of doped graphene as catalyst instead GO are also require molecular hydrogen from outside.[15] The minimal number of reaction steps is crucial for the employment of GO as a catalyst because each step of the reaction is also a step of GO reduction that provides a significant additional decrease of the reaction yield. [16] For the case of carbon monoxide conversion we propose a similar scheme where the second and third steps of the conversion are the same as the third and fourth for the carbon dioxide case:

$CO + 2H^+ + 2e^- \rightarrow HCHO$ (2a)

$HCHO + 2H^+ + 2e^- \rightarrow CH_3OH$ (3a)

$CH_3OH + 2H^+ + 2e- \rightarrow CH_4 + H_2O$ (4a)

The main contribution to the energy cost of the reactions discussed in Ref. [16] comes by the formation of dangling bonds [30] created by the removal of an odd number of hydroxyl groups. For the case of significant oxidation of graphene and high concentration of reduced species this issue plays no role because multiple similar processes can occur simultaneously and an even number of hydroxyl group can transform to epoxy groups at one time. When significant reduction of GO in catalytic process is achieved the problem of a high energy barrier at intermediate steps arises again. Significant acceleration of the nitroarenes hydration after addition to reduced GO gold nanoparticles [38] suggest the route to solve this problem – addition of a second catalytic species which can be a temporary host for the hydrogen atoms from other hydroxyl groups. On the other hand, Mn4-clusters have been considered important for the processes of photosynthesis.[39,40] The possibility of varied oxidation states of manganese ions in these clusters enables these molecules temporarily host hydrogen atoms during the $CO_2$ conversion process.[39,40] Other transition metal molecular systems have been also proposed for $CO_2$ hydration (see Ref. [41] and references therein). Insertion of the large molecules between GO layers and the fabrication of large numbers of composites from transition metals nanoparticles and GO [42-44] enables the use of transitional metal-based nanosystems for the increasing the reaction rate over GO catalysts.

A model of significantly reduced GO as used in our previous work [16] has been employed. This model allows us to estimate the possibility of the discussed processes after a visible level of GO reduction during the catalytic process. [10,11,18] As a model of a metal-oxide nanocluster we used a simple $MO_2$ molecule, where M a V, Cr, Mn, or Fe atom. The choice of the transition metals was guided by the multiplicity of the oxidation states of these metals in oxides, which is crucial for the catalytic properties. This molecule is the minimal model of the molecules like $Mn_4$–clusters [36] and is also feasible for the modeling of the interaction of graphene with oxides-based nanoparticles. [45] To check the possible role of the transition steps (the steps between initial and intermediate and intermediate and final stages of the reaction) we performed the calculation of the energy costs of these steps using the method described in detail in Ref. [44] and used for the modeling of the various reactions over GO substrate, [16] and we find that the energy costs of these steps is much less that the energy cost of the formation of the energetically unfavorable transition stage (about +0.3 eV). Thus the energy cost of this step of the reaction corresponds with the energy difference between initial and intermediate stages (Fig. 1 a,b).

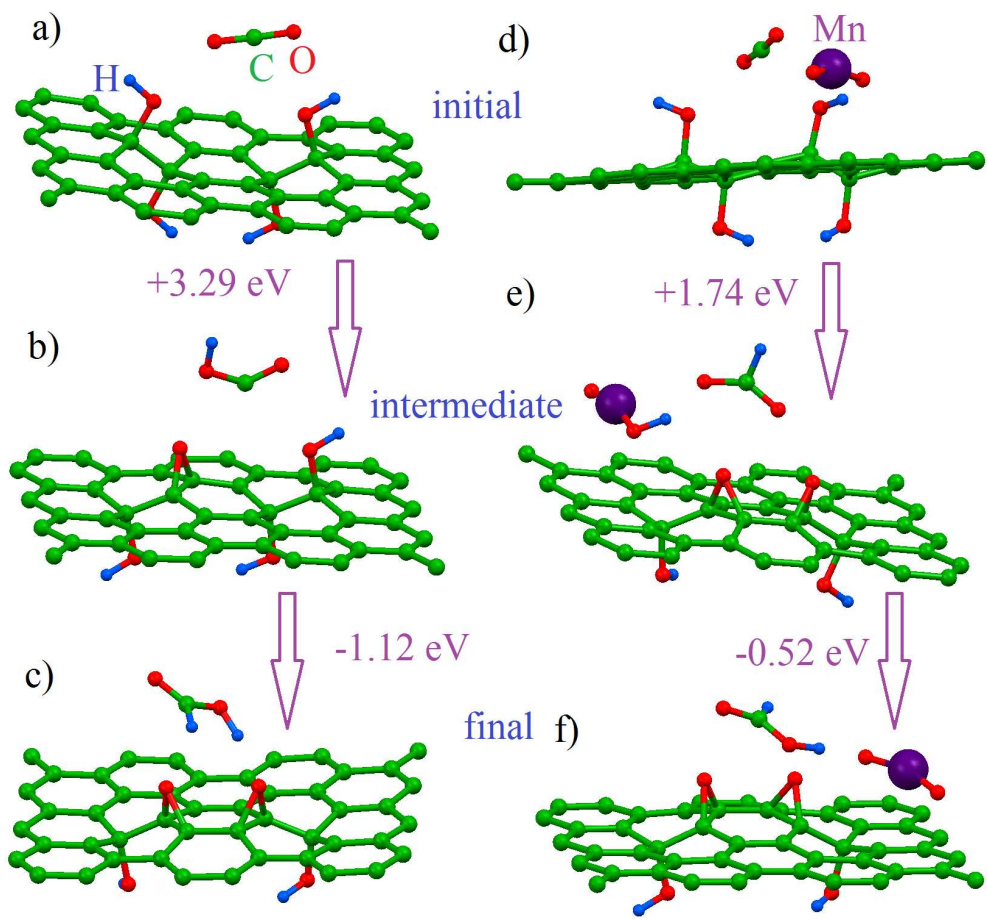

**Figure 1** Optimized atomic structure and energy costs of initial (a, d), intermediate (b, e) and final (c, f) stages of $CO_2$ to HCOOH conversation (first step of $CO_2$ to $CH_4$ conversation) over pure graphene oxide (a-c) and mix of graphene oxide and metal oxide (d-f).

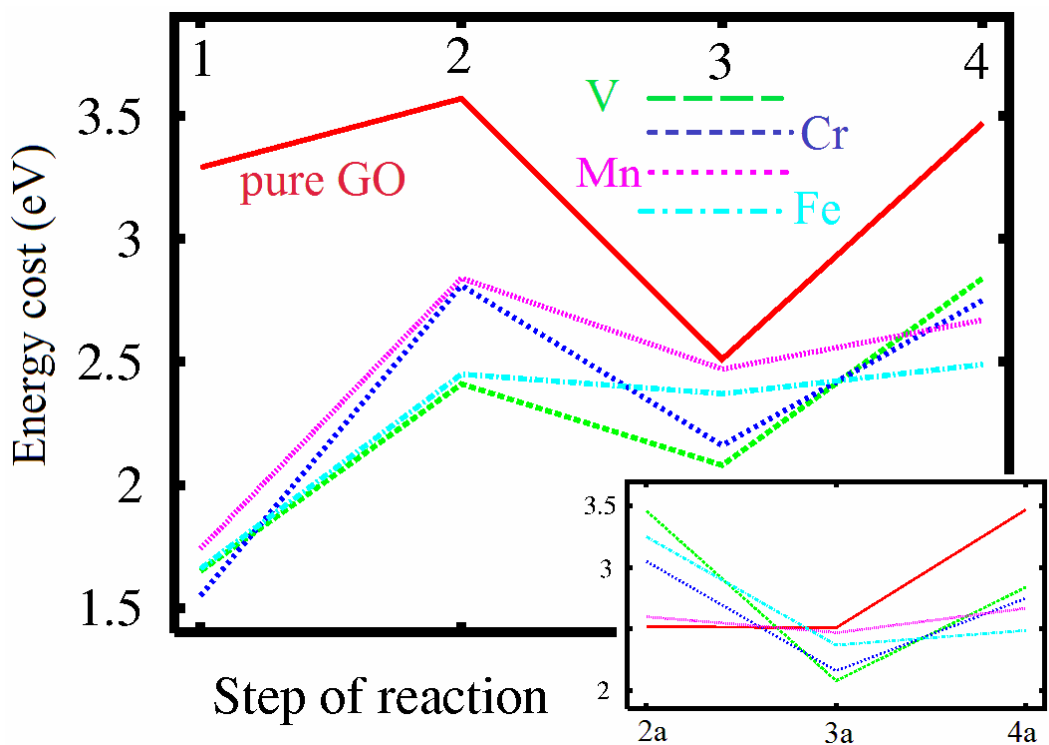

**Figure 2** Energy costs of each of the steps of conversion (see description in text) of carbon mono (inset) and dioxide to methane over the pure and metal oxide decorated graphene oxide.

**Hydration of carbon dioxide**

The energy cost of the reaction is defined as the difference between the total energies of the system before and after each stage of reaction. The energy cost of each step of the reaction is defined as maximal value of the energy costs of each step of reactions. For example, reaction 1 over GO scaffold proceeds in two stages. In the first stage hydrogen from the nearest hydroxyl group on GO surface migrates to the $CO_2$ molecule and transforms it to COOH. Residual oxygen from the hydroxyl group then transforms to the epoxy group (see Fig. 1a,b). The intermediate stage of the reaction leads to the formation of the two radical like structures. One is the COOH molecule, and the second is GO with an odd number of hydroxyl groups. The presence of the even number of covalently bonded species (hydrogen, fluorine, hydroxyl groups) leads to the formation of an energetically unfavorable and metastable configuration. [30,31] The formation of these two radical-like structures provides the higher energy cost of this stage of the first step of the reaction (+3.29 eV). Similar to other radical–like structures discussed in our recent works, [16,30,31] this intermediate configuration is also unstable and requires migration of the hydrogen atom from the other hydroxyl group in vicinity to the COOH molecule with formation of the neutral HCOOH. The number of hydroxyl groups on GO became even and there are no more unpaired electrons in graphene. This transformation of intermediate to the final configuration is energetically very favorable (-1.12 eV).

The modeling of the other (2-4) steps of the transformation of $CO_2$ to methane has been explored within this scheme. Results of calculations of the energetics of these steps are reported in Fig. 2. We can see this reaction has rather higher energy costs. To estimate reaction yields at room temperature for the studied reaction we use the standard formula:

$C = C_0 e^{-E/k_BT}$,

where E is the energy cost of the step of reaction, T - temperature, $k_B$ - Boltzmann constant, $C_0$ – coefficient for this type reaction and temperatures and C – reaction yield. The $C_0$ coefficient for the room temperature can be calculated from the experimentally observed GO re-oxidation at room temperature. The calculated value of the reaction yield for the first step of $CO_2$ to $CH_4$ transformation is 0.015%. The total yield of the reaction depends on the maximal value of the energy costs of the steps and for the studied reaction over pure GO it is 0.01%. This value is of the same order as the efficiency of $CO_2$ hydration over platinum and ruthenium-based photocatalysts. To check the role of the level of reduction of GO for the energetics of the considered processes we performed the calculations for the case of unreduced GO (75% of carbon atoms covalently bonded with epoxy and hydroxyl groups, as employed in our previous work) [18] and find that similar to the case of the oxidation of various organic species, and increase of the oxidation level of GO increases the energy costs of reaction by about 0.4 eV. The obtained values of the energy costs provide evidence for the weak dependence of the catalytic activity of GO from the level of reduction, in agreement with experimental results. [10,11]

**Increasing of reaction rate in the presence of transitional metal oxides**

Addition of a $MnO_2$ molecule to the system (Fig. 1d) significantly changes the pathway and energetics of the reaction. On the first stage of the first step of reaction (Fig. 1d,e) a hydrogen atom from one hydroxyl group transforms $CO_2$ to HCOO (instead of COOH as in the case of pure GO) and hydrogen from the second hydroxyl group moves to the oxygen atom of the metal oxide and transforms $Mn^{IV}O_2$ to $Mn^{III}OOH$. Multivalency of several transition metals permits oxides of these metals to host the hydrogen atoms and in our case avoids the formation of the radical–like energetically unfavourable structure with the odd number of hydroxyl groups on the graphene sheet. Changes of the reaction pathways in the presence of $MO_2$ molecules lead to significant decreases of the energy costs of each step of the reaction (see Fig. 1d-f and Fig. 2). The best transition metals for the studied reactions are iron and vanadium. The maximal energy cost of the step of $CO_2$ hydration does not exceed 2.5 eV, corresponding to a yield at room temperature about 0.07% which is higher than for the case of pure GO. Increasing the temperature of reaction from room temperature to 100 °C will increase reaction yields to 100%.

**Hydration of carbon monoxide**

In the case of CO to $CH_4$ conversion the last two steps of the reactions are the same, and only the initial step of the reaction (2a) is significantly different from the case of $CO_2$ hydrogenation. The energy cost of this step is surprisingly low for the case of pure GO and for the cases with the presence of metal oxides (except manganese). This effect is caused by the spontaneous migration of the second hydrogen atom from the hydroxyl group to the CO molecule yielding the formation of HCHO without formation of the intermediate structures. But for the last two steps of reaction the presence of metal oxides is required for decreasing the energy costs. Thus, in the graphene oxide/metal oxide composites step 2a will involve only GO, and the last steps will employ the metal oxide sites. The highest energy cost of carbon monoxide conversion is the same as in $CO_2$ case – about 2.5 eV which corresponds to a reaction yield at room temperature of about 0.07%.

**Conclusions**

First principles density functional theory-based modeling demonstrates possibility of the carbon mono and dioxide conversion to methane via the employment of graphene oxide as source of the hydrogen atoms for the hydrogenation process. The transformation of hydroxyl groups on the graphene surface to epoxy groups is reversible at room temperature in the presence of the water. [10,16] The presence of the transition metal oxides significantly decreases the energy costs of the reactions. This decreasing requires a further detailed exploration toward the most efficient metal oxide-based part of the catalyst to obtain a maximal reaction rate. Estimated yields of carbon mono and dioxide conversion to methane are 0.01% for pure graphene oxide and 0.07% for the case of graphene oxide/metal oxide composites, and there is total conversion of both gases at temperatures of 100 °C. Obtained reaction rates are the same order as for the discussed before platinum and ruthenium-based photocatalysts [20] but the lighter weight and larger surface area makes the reaction yield per gram or $cm^2$ of graphene oxide much higher. Adding photoactive centers to the system [40] could also be a source of increasing of the conversion rate. Reported results demonstrate that graphene oxide/metal oxide composites are attractive not only as storage materials [41] but also for catalysis of various reactions at low temperatures.

**Acknowledgements**

I acknowledge computational support from the CAC of KIAS and dr. R. Green for the careful reading of the manuscript.